# Superconductivity in Hexagonal $Zr_6CoAl_2$-Type $Zr_6RuBi_2$ and $Zr_6FeBi_2$

Kosuke Yuchi*, Daisuke Nishio-Hamane, Keita Kojima, Kodai Moriyama,
Ryutaro Okuma, and Yoshihiko Okamoto†

*Institute for Solid State Physics, University of Tokyo, Kashiwa 277-8581, Japan*

We report the synthesis and electronic properties of polycrystalline samples of $Zr_6MBi_2$ (M = Ru and Fe) crystallizing in the hexagonal $Zr_6CoAl_2$-type structure. Based on their electrical resistivity, magnetization, and heat capacity data, $Zr_6RuBi_2$ and $Zr_6FeBi_2$ are found to exhibit bulk superconductivity below $T_c$ = 4.9 and 1.4 K, respectively. Although $Zr_6RuBi_2$ is most likely a conventional superconductor, the considerably higher $T_c$ for M = Ru than that for M = Fe differs from the trend in $T_c$ for $Zr_6CoAl_2$-type superconductors reported thus far. The superconductivity of an amorphous solid-solution phase, which may hinder elucidation of the superconducting properties of $Zr_6MBi_2$, is also discussed.

Recently, new superconductors have been discovered in $A_6MX_2$ compounds with the hexagonal $Zr_6CoAl_2$-type crystal structure. Figure 1(a) shows this crystal structure, which is an ordered $Fe_2P$ type with the noncentrosymmetric space group *P*-62*m*. The discovery of superconductors in the $A_6MX_2$ family was triggered by $Sc_6MTe_2$, which was found to exhibit superconductivity in the cases of seven transition metals M (M = Fe, Co, Ni, Ru, Rh, Os, and Ir) [1]. The superconducting transition temperature $T_c$ is approximately 2 K for M = 4*d* and 5*d* transition metals, whereas in the M = 3*d* cases, $T_c$ increases in the order of Ni, Co, and Fe with the highest value of $T_c$ = 4.7 K being observed for $Sc_6FeTe_2$. Subsequently, $Zr_6FeTe_2$, $Zr_6CoTe_2$, and $Zr_6FeSb_2$ were found to exhibit superconductivity at lower temperatures than $Sc_6MTe_2$ [2,3]. According to the results of first-principles calculations, electronic states near the Fermi level in $Sc_6MTe_2$ and $Zr_6MTe_2$ mainly consist of *d* orbitals of M and Sc/Zr atoms, and especially in $Sc_6FeTe_2$, which shows the highest $T_c$, the contribution of the Fe 3*d* orbitals is considerably large [1,2,4]. $Sc_6FeTe_2$ has also been shown to have a strong lattice anharmonicity compared to other $Sc_6MTe_2$ due to the rattling of Fe atoms [4], which may be related to the high $T_c$ of this material. In contrast, $Sc_6MnTe_2$ and $Zr_6MnTe_2$ do not exhibit superconductivity, but show strong magnetism originating from the Mn 3*d* electrons [1,2]. Thus, $Zr_6CoAl_2$-type $A_6MX_2$ constitute a *d*-electron superconductor family, in which the *d* electrons of the M and A atoms are strongly involved in the emergence of superconductivity, making them a potential target for studying the correlation between superconductivity and various features of transition metals.

As a superconductor family, a remarkable feature of $A_6MX_2$ is that the A, M, and X atoms can all be changed to various elements. For example, both $Sc_6NiTe_2$ and $Zr_6FeSb_2$ exhibit superconductivity. In $Zr_6CoAl_2$-type $A_6MX_2$, the A atoms can be Sc, Y, Zr, Hf, or the heavy lanthanides, the M atoms can be almost all transition metals, and the X atoms can be various electronegative elements of groups 13 to 16. The number of the permutations of these A, M, and X atoms is very large, and over 100 materials have been synthesized to date [1-3, 5-25]. There have been many studies on the physical properties of $A_6MX_2$ compounds, including on the spin structure of the magnetically ordered phase [22, 25-29] and their magnetocaloric properties [30-32]. However, to the best of our knowledge, there are no reports on superconductivity other than on the above ten materials [1-3] and many $A_6MX_2$ compounds have yet to be characterized, raising expectation that many *d*-electron superconductors will be discovered in this family.

In this letter, we focus on $Zr_6MBi_2$ (M = Ru and Fe) crystallizing in the $Zr_6CoAl_2$-type structure. For M = Mn, Fe, Co, Ni, Cu, and Rh, polycrystalline samples have been synthesized, and the crystal structure was reported to be of the $Zr_6CoAl_2$ type [11, 20, 25]. Polycrystalline samples of $Zr_6FeBi_2$ and $Zr_6CoBi_2$ were found to show metallic electrical resistivity above 4.2 K [11]. No superconductivity has been reported, even when extending the range to all $A_6MX_2$ members with X = Bi. New superconductors with X = Bi will provide an insight into heavy element effects of the X atoms on superconductivity. In the case of $Sc_6MTe_2$, M = Os and Ir exhibit lower $T_c$ values than those of M = Fe and Co, but considerably higher upper critical fields $H_{c2}$ for their $T_c$ values, suggesting that strong spin-orbit coupling of the 5*d* electrons of M atoms plays a role [1].

Polycrystalline samples of $Zr_6MBi_2$ (M = Ru and Fe) were synthesized by the arc-melting of Zr chips (99.9%, RARE METALLIC), Ru (99.95%, RARE METALLIC) or Fe (99.5%, RARE METALLIC) powder, and Bi chips (99.99%, Wako Pure Chemical Corp.). First, 6:1.5:2.5 and 6:1:2.5 molar ratios of Zr chips, M powder, and Bi chips for M = Ru and Fe were respectively placed on a water-cooled copper hearth and arc-melted under an Ar atmosphere. The obtained buttons were subsequently inverted and arc-melted several times to promote homogenization. They were then annealed in evacuated quartz tubes at 1273 K for 216 h. For $Zr_6RuBi_2$, the sample prepared by this procedure is named as sample #1. Hereinafter, mainly the experimental data for this sample are presented, and those of the sample #2, prepared under different conditions are only presented in the last part of this letter. Unless specified otherwise, we refer to sample #1. For $Zr_6FeBi_2$, all of the experimental data presented herein were obtained using this sample. Microstructural observations and chemical analyses were conducted by means of a scanning electron microscope (SEM; JEOL IT-100) equipped with an attachment for energy-dispersive X-ray spectroscopy (EDX; 15 kV, 0.8 nA, 1 μm beam diameter). The ZAF method was used for data correction, and the standards used were samples of the pure metals. Powder X-ray diffraction (XRD) measurements on $Zr_6RuBi_2$ (sample #1) and $Zr_6FeBi_2$ were performed on a Debye- Scherrer diffractometer R-AXIS RAPID II (RIGAKU) using Mo K$\alpha$ radiation at room temperature. Details on the measurements are described in Supplementary Note 2. The structural analysis was performed using JANA2006 [33]. A powder XRD measurement on $Zr_6RuBi_2$ (sample #2) was performed on a Bragg-Brentano diffractometer MiniFlex600-C (RIGAKU) using Cu K$\alpha$ radiation at room temperature. Electrical resistivity and heat capacity measurements were performed using a Physical Property Measurement System (Quantum Design). Magnetization measurements were performed using an MPMS-3 (Quantum Design). Electrical resistivity measurements down to 0.1 K were performed using an adiabatic demagnetization refrigerator. Heat capacity and magnetization measurements down to 0.5 K were performed using a $^3$He refrigerator.

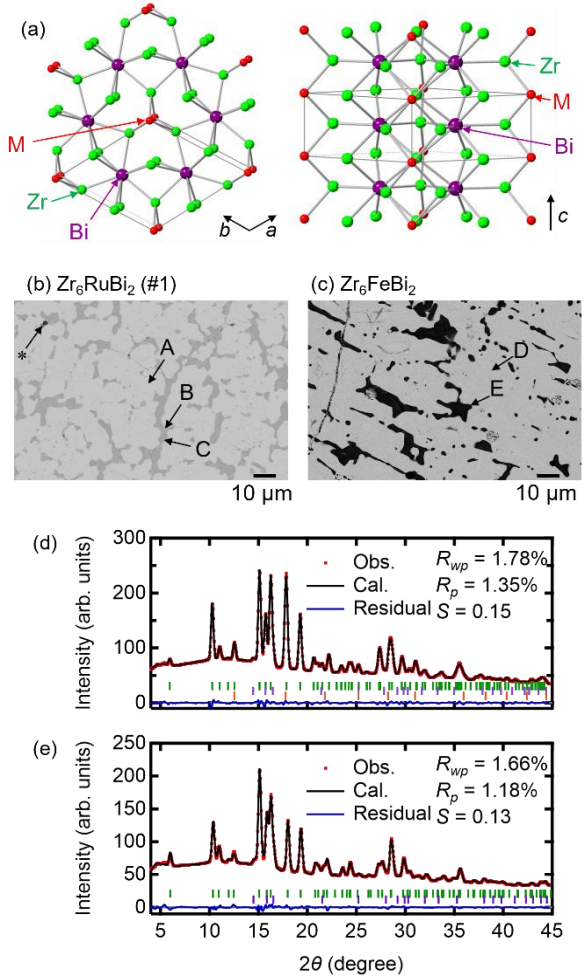

Figure 1. (a) Crystal structure of $Zr_6MBi_2$ (M = Ru and Fe) viewed along (left) and perpendicular to (right) the $c$ axis. The solid lines indicate the unit cell. (b, c) SEM images of (b) $Zr_6RuBi_2$ (sample #1) and (c) $Zr_6FeBi_2$ polycrystalline samples. In (b), the black area marked with an asterisk is a void. (d, e) Powder XRD patterns of (d) $Zr_6RuBi_2$ (sample #1) and (e) $Zr_6FeBi_2$ polycrystalline samples acquired at room temperature using R-AIXS RAPID-II (Mo K$\alpha$). Filled circles show experimental data. The overplotted curve shows the calculated pattern, whereas the lower curve shows a difference plot between the experimental and calculated intensities. The top (green), second (purple), and third (orange, $Zr_6RuBi_2$ only) vertical bars indicate the position of the Bragg reflections of the $Zr_6MBi_2$, Zr, and ZrRu phases, respectively. The $S$ values are small caused by high background intensity due to X-ray fluorescence.

Figures 1(b) and 1(c) show SEM images of polycrystalline samples of $Zr_6RuBi_2$ (sample #1) and $Zr_6FeBi_2$, respectively. In the SEM image of $Zr_6RuBi_2$, the white area marked A is the main phase, which is accompanied by two minor phases marked B and C (B and C areas are both gray, but area B is slightly paler than area C. Figures that allow the difference between B and C areas to be more easily recognized are shown in Supplementary Fig. 1). The percentages of phases A, B, and C of the total area are 81%, 9%, and 9%, respectively. The chemical compositions of phases A, B, and C were determined by EDX as $Zr_{6.06(2)}Ru_{1.00(1)}Bi_{1.94(1)}$, $Zr_{0.51}Ru_{0.49}$, and $Zr_{0.99}Bi_{0.01}$, respectively, where the total number of atoms is fixed to be nine for phase A and one for phases B and C. The results indicate that all three phases are almost stoichiometric. Considering the small degree of nonstoichiometry and powder XRD results described below, the chemical formulae of phases A, B, and C are expressed as $Zr_6RuBi_2$, ZrRu, and Zr, respectively. Among these phases, ZrRu and Zr do not show superconductivity. In the SEM image of $Zr_6FeBi_2$ shown in Fig. 1(c), the gray region marked D is the main phase and the black region marked E is a minor phase. The percentages of the phases D and E of the total area are 89% and 11%, respectively. The EDX chemical analysis as applied to $Zr_6RuBi_2$ showed the chemical compositions of phases D and E to be $Zr_{6.07(1)}Fe_{0.98(2)}Bi_{1.95(1)}$ and $Zr_{0.987}Fe_{0.007}Bi_{0.006}$, respectively. As in the case of $Zr_6RuBi_2$, considering the small degree of nonstoichiometry and powder XRD results, the chemical formulae of these phases are expressed as $Zr_6FeBi_2$ and Zr, respectively. The Zr phase does not exhibit super-conductivity. These SEM and EDX results clearly show that each sample contains $Zr_6MBi_2$ as the main phase, accompanied by nonsuperconducting minor phases.

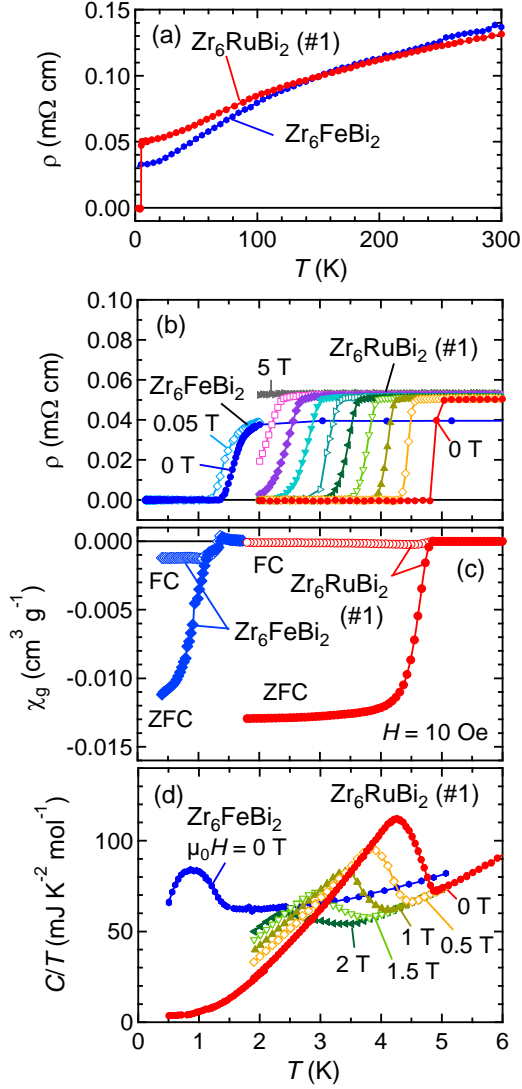

Figure 2. Electronic properties of polycrystalline samples of $Zr_6RuBi_2$ (sample #1) and $Zr_6FeBi_2$. (a) Temperature dependence of electrical resistivity of $Zr_6RuBi_2$ and $Zr_6FeBi_2$ measured above 2 K. (b) Temperature dependence of electrical resistivity of $Zr_6RuBi_2$ at $\mu_0 H = 0, 0.5, 1, 1.5, 2, 2.5, 3, 3.5, 4$, and 5 T and of $Zr_6FeBi_2$ at $\mu_0 H = 0$ and 0.05 T. (c) Temperature dependence of FC and ZFC magnetic susceptibility of $Zr_6RuBi_2$ and $Zr_6FeBi_2$ measured at a magnetic field of 10 Oe. (d) Temperature dependence of heat capacity divided by temperature for $Zr_6RuBi_2$ measured at various magnetic fields and $Zr_6FeBi_2$ at zero magnetic field. The $C/T$ data were calculated assuming each sample to consist of a single phase of $Zr_6MBi_2$.

Figures 1(d) and 1(e) shows the powder XRD patterns of $Zr_6RuBi_2$ (sample #1) and $Zr_6FeBi_2$ polycrystalline samples, respectively, along with the results of their Rietveld analyses. The crystallographic parameters determined from these analyses are listed in Supplementary Table I. Consistent with the SEM and EDX results, diffraction peaks of Zr are present in the XRD patterns of both samples and those of ZrRu are present in that of $Zr_6RuBi_2$. Therefore, we performed multiphase analyses including these impurity phases. In each case, the difference between experimental and calculated intensities was minimized when a hexagonal $Zr_6CoAl_2$-type structural model was used for the main phase, indicating that $Zr_6CoAl_2$-type $Zr_6MBi_2$ was obtained as the main phase for both M = Ru and Fe. The lattice parameters were refined to $a$ = 7.9181(3) Å and $c$ = 3.6847(2) Å for M = Ru and $a$ = 7.8629(4) Å and $c$ = 3.7174(3) Å for M = Fe. To the best of our knowledge, $Zr_6RuBi_2$ has not been reported previously. However, the $a$ and $c$ values of $Zr_6FeBi_2$ are consistent with those of a previous study [11].

Figures 2(a) and 2(b) show the temperature dependences of electrical resistivity, $\rho$, of the $Zr_6RuBi_2$ (sample #1) and $Zr_6FeBi_2$ polycrystalline samples. As shown in Fig. 2(a), both samples showed metallic $\rho$, which decreased with decreasing temperature. The residual resistivity ratios, RRR = $\rho_{300K}/\rho_0$, where $\rho_{300K}$ is $\rho$ at 300 K and $\rho_0$ is the residual resistivity, were evaluated as 2.6 and 4.5 for $Zr_6RuBi_2$ and $Zr_6FeBi_2$, respectively. At lower temperatures, the $\rho$ of both samples showed a sharp drop to zero. The onset and zero-resistivity temperatures were determined as 5.0 and 4.8 K for $Zr_6RuBi_2$ and 2.0 and 1.4 K for $Zr_6FeBi_2$, respectively.

As shown in Fig. 2(c), the zero-field-cooled (ZFC) and field-cooled (FC) magnetization data of $Zr_6RuBi_2$ (sample #1) and $Zr_6FeBi_2$ polycrystalline samples showed strong diamagnetic signals due to superconductivity below 4.9 and 1.4 K, respectively. The shielding fractions of $Zr_6RuBi_2$ (1.8 K) and $Zr_6FeBi_2$ (0.4 K), evaluated by assuming a single phase of $Zr_6MBi_2$, were determined as 144% and 120%, respectively. Furthermore, as shown in Fig. 2(d), the values of heat capacity divided by temperature, $C/T$, of the polycrystalline samples of $Zr_6RuBi_2$ (sample #1) and $Zr_6RuBi_2$ strongly increased from 4.9 and 1.4 K and reached maxima at 4.2 and 0.9 K, respectively, with decreasing temperature. These results clearly indicate that both samples exhibited bulk superconductivity. Considering the midpoint of the resistivity drop, the onset of the magnetization drop, and the onset of the heat capacity jump, $T_c$ was determined as 4.9 and 1.4 K for M = Ru and Fe, respectively.

In the following, we discuss superconducting properties on the basis of heat capacity data. As shown in Fig. 3(a), normal-state $C/T$ for both M = Ru and Fe are linear with respect to $T^2$ at low temperatures. The solid lines show the results of fitting to the equation $C/T = AT^2 + \gamma$, where $A$ and $\gamma$ represent the coefficient of the $T^3$ term of the lattice heat capacity and Sommerfeld coefficient, respectively. The $T^3$ term and Sommerfeld coefficient for $Zr_6RuBi_2$ and $Zr_6FeBi_2$ were estimated as $A$ = 1.563(9) and 0.72(1) mJ K$^{-4}$ mol$^{-1}$ and $\gamma$ = 34.7(1) and 59.37(9) mJ K$^{-2}$ mol$^{-1}$, respectively. Temperature dependence of the electronic heat capacity divided by temperature, $C_{el}/T$, obtained by subtracting the lattice contribution from experimental data, i.e., $C_{el}/T = C/T − AT^2$, is plotted in Fig. 3(b). The $C_{el}/T$ data of $Zr_6RuBi_2$ below $0.35T_c$ were fitted to the equation $C_{el} = \gamma_0 T + B\exp[-\Delta(0)/k_BT]$, where $\Delta(0)$ is the size of the superconducting gap at zero temperature. The result of this fitting is reasonably good, suggesting the presence of an isotropic gap and yielding $\gamma_0 = 3.20(6)$ mJ $K^{-2}$ $mol^{-1}$, $B = 1.3(1) \times 10^3$ mJ $K^{-1}$ $mol^{-1}$, and $\Delta(0)/k_B = 7.4(1)$ K. The obtained $\gamma_0$ is 9% of $\gamma$, suggesting that the majority of the sample is superconducting at the lowest temperature.

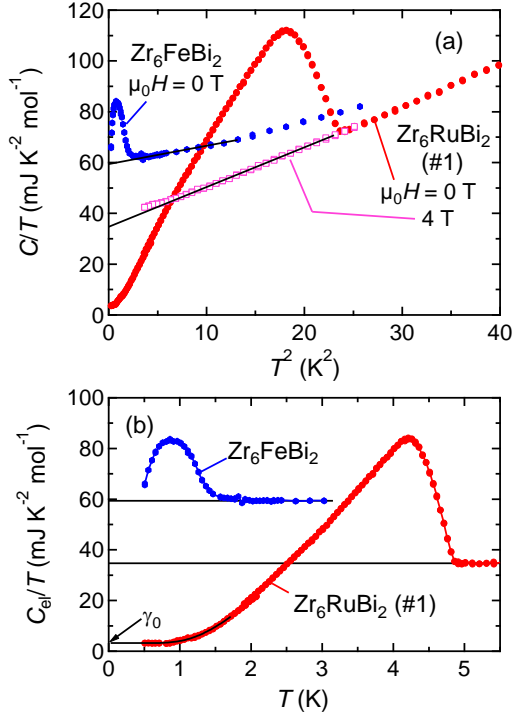

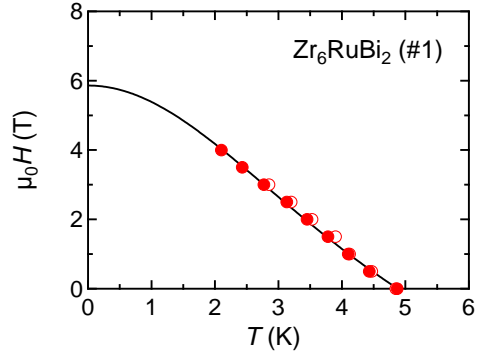

Figure 4. Magnetic field versus midpoint temperatures in the resistivity drop (filled circles) and onset temperatures in the heat capacity jump (open circles) for polycrystalline $Zr_6RuBi_2$ (sample #1). The solid curve shows a fitting to the Ginzburg–Landau formula.

Figure 3. (a) Heat capacity divided by temperature of $Zr_6RuBi_2$ (sample #1) and $Zr_6FeBi_2$ polycrystalline samples as a function of $T^2$, calculated assuming a single phase of $Zr_6MBi_2$. The solid lines show the results of linear fits of the 4 T data for $Zr_6RuBi_2$ between 3 and 4.8 K and zero-field data for $Zr_6FeBi_2$ between 2 and 3.6 K. (b) Temperature dependence of electronic heat capacity divided by temperature for $Zr_6RuBi_2$ and $Zr_6FeBi_2$ obtained by subtracting the lattice contribution from the $C/T$ data. The solid lines indicate the normal state $\gamma$ for each compound. The solid curve on the $Zr_6RuBi_2$ data shows a fitting to the equation $C_{el} = \gamma_0 T + B\exp[-\Delta(0)/k_BT]$.

Using the heat capacity jump at $T_c$ of $\Delta C_{el}/T_c = 49.5$ mJ $K^{-2}$ $mol^{-1}$ and the superconducting contribution of the Sommerfeld coefficient $\gamma_s = \gamma - \gamma_0 = 31.5$ mJ $K^{-2}$ $mol^{-1}$, the magnitude of the jump in $Zr_6RuTe_2$ was estimated as $\Delta C_{el}/\gamma_s T_c = 1.6$. This value is somewhat underestimated by a gradual change of $C_{el}/T$ at $T_c$, suggesting that the actual $\Delta C_{el}/\gamma_s T_c$ is slightly larger than the weak-coupling limit value of 1.43 for a conventional Bardeen–Cooper-Schrieffer (BCS) superconductor. This result suggests that the superconductivity in $Zr_6RuBi_2$ is most likely a strong-coupling BCS superconductivity, which would be consistent with the observation that the low-temperature $C_{el}$ is well-fitted by the exponential function, as shown in Fig. 3(b). In contrast, $\gamma_0$ of $Zr_6FeBi_2$ could not be determined due to the low $T_c$. Moreover, the broad jump in $C_{el}/T$, as shown in Fig. 3(b), does not allow an accurate estimate of $\Delta C_{el}$. In order to discuss the superconducting properties of $Zr_6FeBi_2$ based on the heat capacity data, it would be necessary to perform experiments at lower temperatures using a sample of higher quality.

The superconducting properties of $Zr_6RuBi_2$ under magnetic fields can be discussed on the basis of the $\rho$ and $C/T$ data measured in various magnetic fields, as shown in Figs. 2(b) and 2(d), respectively. As shown in Fig. 4, the midpoint of the decrease in $\rho$, $T_c^{mid}$, and the onset of $T_c$ of heat capacity jump show similar trends in magnetic fields, indicating that $T_c^{mid}$ reflects the bulk superconducting transition. Fitting $H_{c2}(T)$ estimated from $T_c^{mid}$ with the Ginzburg–Landau (GL) formula, $H_{c2}(T) = H_{c2}(0)[1 - (T/T_c^{mid})^2]/[1 + (T/T_c^{mid})^2]$, yields $\mu_0H_{c2}(0) = 5.86(5)$ T and GL coherence length of $\xi = 0.75$ nm. This $H_{c2}(0)$ is considerably lower than the Pauli-limiting field of $\mu_0H_{c2}(0)/T = 1.84T_c^{mid}/K = 9$ T [34], suggesting that the strong spin-orbit coupling of heavy elements does not significantly affect the superconducting properties of $Zr_6RuBi_2$ under magnetic fields, even though it has a noncentrosymmetric crystal structure and comprises heavy elements.

As discussed above, superconductivity in $Zr_6RuBi_2$ is most likely conventional, but it is noteworthy that its $T_c$ is considerably higher than that of $Zr_6FeBi_2$. We synthesized $Zr_6MBi_2$ samples with other M atoms such as Rh, Ir, and Co. However, no material other than M = Ru was confirmed to exhibit bulk superconductivity above 2 K, meaning that an anomalously high $T_c$ is manifested in $Zr_6RuBi_2$. This trend in $T_c$ differs from that for $Sc_6MTe_2$, where $Sc_6FeTe_2$ showed the highest $T_c$ of 4.7 K [1]. $Zr_6FeTe_2$ also showed the highest $T_c$ among $Zr_6MTe_2$, although its $T_c$ was much lower [2]. The fact that $Zr_6RuBi_2$ showed the highest $T_c$ among the current $A_6MX_2$ members indicates that M being a 3d transition metal element is not always advantageous for superconductivity, and that the combination of A, M, and X elements is important. In particular, $\gamma$ = 36.3 mJ K$^{-2}$ mol$^{-1}$ for Zr$_6$RuBi$_2$ is considerably smaller than $\gamma$ = 73 mJ K$^{-2}$ mol$^{-1}$ for Sc$_6$FeTe$_2$ ($T_c$ = 4.7 K) and $\gamma$ = 50 mJ K$^{-2}$ mol$^{-1}$ for Zr$_6$FeBi$_2$ ($T_c$ = 1.4 K), but is comparable to $\gamma$ = 40 mJ K$^{-2}$ mol$^{-1}$ for Sc$_6$RuTe$_2$ ($T_c$ = 1.9 K) [1]. It is surprising that the highest $T_c$ in the A$_6$MX$_2$ family is achieved in Zr$_6$RuBi$_2$. Future detailed elaboration of the trend in $T_c$ in Zr$_6$MBi$_2$ based on the experimental results of other M cases and electronic structure calculation results will allow us to pursue the realization of a higher $T_c$ in A$_6$MX$_2$.

Finally, we discuss a difficulty regarding the study of superconductivity in Zr$_6$MBi$_2$ from a chemical point of view. Supplementary Fig. 2(a) shows a SEM image of a Zr$_6$RuBi$_2$ sample prepared by arc-melting of stoichiometric amounts of Zr, Ru, and Bi without annealing. This sample is hereinafter referred to as sample #2. As shown in this figure, sample #2 consisted of two different phases marked F (light-gray) and G (dark-gray). According to EDX analysis, phase F is the Zr$_6$RuBi$_2$ phase with the chemical composition Zr$_{6.03(4)}$Ru$_{0.82(3)}$Bi$_{2.15(3)}$, whereas phase G is a Zr-Ru-Bi solid solution with composition Zr$_{0.90}$Ru$_{0.05}$Bi$_{0.05}$. These Zr$_6$RuBi$_2$ and Zr-Ru-Bi solid-solution phases account for 61% and 39% of the total area, respectively, indicating that while Zr$_6$RuBi$_2$ is the main phase, the Zr-Ru-Bi solid solution is also present in a significant proportion. However, as shown in Supplementary Fig. 2(b), only the diffraction peaks of Zr$_6$RuBi$_2$ with the Zr$_6$RuBi$_2$-type crystal structure appear in the powder XRD pattern of sample #2, indicating that the Zr-Ru-Bi solid-solution phase was amorphous. This amorphous phase could not be completely eliminated by annealing sample #2.

In sample #2, not only the Zr$_6$RuBi$_2$ phase but also the Zr-Ru-Bi solid solution exhibit superconductivity at a similar $T_c$. As shown by the $C/T$ data in Supplementary Fig. 3(a), in addition to the peak at $T_{c1}$ = 4.3 K, there is a shoulder-like feature at around $T_{c2}$ = 4.6 K. As shown in Supplementary Fig. 3(b), the $T_{c1}$ of sample #2 shows a similar variation under magnetic fields as that of sample #1, whereas the change in $T_{c2}$ by applying a magnetic field is smaller than that of $T_{c1}$, suggesting that $T_{c1}$ and $T_{c2}$ correspond to the superconducting transitions of the Zr$_6$RuBi$_2$ and amorphous Zr-Ru-Bi phases, respectively. Amorphous phases comprising Zr and several percent of a transition metal element have been reported to exhibit superconductivity at $T_c$ = 4–6 K [35], supporting the above consequence. The superconducting transition at $T_{c2}$ appeared in Zr$_6$RuBi$_2$ samples synthesized under various conditions except for sample #1. In sample #1 of Zr$_6$RuBi$_2$, the nominal chemical composition was shifted from the stoichiometric ratio and the sample was annealed under the appropriate conditions, resulting in suppression of formation of the amorphous superconducting phase. This amorphous superconducting phase hinders the exploration of new Zr$_6$MBi$_2$ superconductors. It is expected that new superconductors will be discovered in the A$_6$MX$_2$ family in future studies, but our results for sample #2 suggest that in this quest the possible formation of an amorphous impurity phase should always be borne in mind.

In summary, Zr$_6$CoAl$_2$-type Zr$_6$RuBi$_2$ and Zr$_6$FeBi$_2$ have been identified as bulk superconductors with $T_c$ = 4.9 K and 1.4 K, respectively, based on electrical resistivity, magnetization, and heat capacity measurements on polycrystalline samples. The $T_c$ of Zr$_6$RuBi$_2$ is highest in the A$_6$MX$_2$ family. The superconductivity of Zr$_6$RuBi$_2$ is most likely a conventional BCS superconductivity both in a zero magnetic field and under magnetic fields. However, the considerably higher $T_c$ for Zr$_6$RuBi$_2$ than that for Zr$_6$FeBi$_2$ differs from the trend for known A$_6$MX$_2$ superconductors. We hope that the exploration of new superconductors in the A$_6$MX$_2$ family will lead to the discovery of new superconductors with higher $T_c$ or unusual properties. However, the possible formation of an amorphous superconducting phase should always be borne in mind, since this would hinder elucidation of the superconducting properties of A$_6$MX$_2$.

**Acknowledgments**

The authors are grateful to H. Matsumoto, J. Yamaura, Z. Hiroi, and Y. Shinoda for helpful discussions. This work was supported by JSPS KAKENHI (Grant Nos. 23H01831 and 23K26524) and JST ASPIRE (Grant Number: JPMJAP2314).

**References**

[1] Y. Shinoda, Y. Okamoto, Y. Yamakawa, H. Matsumoto, D. Hirai, and K. Takenaka, J. Phys. Soc. Jpn. **92**, 103701 (2023).
[2] H. Matsumoto, Y. Yamakawa, R. Okuma, D. Nishio-Hamane, and Y. Okamoto, J. Phys. Soc. Jpn. **93**, 023705 (2024).
[3] R. Matsumoto, E. Murakami, R. Oishi, S. Ramakrishnan, A. Ikeda, S. Yonezawa, T. Takabatake, T. Onimaru, and M. Nohara, J. Phys. Soc. Jpn. **93**, 065001 (2024).
[4] M.-C. Jiang, R. Masuki, G.-Y. Guo, and R. Arita, arXiv:2405.10524.
[5] Y.-U. Kwon, S. C. Sevov, and J. D. Corbett, Chem. Mater. **2**, 550 (1990).
[6] F. Gingle, K. Yvon, I. Y. Zavaliy, V. A. Yartys', and P. Fischer, J. Alloys Comp. **226**, 1 (1995).
[7] C. Wang and T. Hughbanks, Inorg. Chem. **35**, 6987 (1996).
[8] H. Kleinke, J. Alloys Comp. **252**, L29 (1997).
[9] H. Kleinke, J. Alloys Comp. **270**, 136 (1998).
[10] I. Y. Zavaliy, V. K. Pecharsky, G. J. Miller, L. G. Akselrud, J. Alloys Comp. **283**, 106 (1999).
[11] G. Melnyk, E. Bauer, P. Rogl, R. Skolozdra, E. Seidl, J. Alloys Comp. **296**, 235 (2000).
[12] P. A. Maggard and J. D. Corbett, Inorg. Chem. **39**, 4143 (2000).
[13] N. Bestaoui, P. S. Herle, and J. D. Corbett, J. Solid State Chem. **155**, 9 (2000).
[14] F. Meng, C. Magliocchi, and T. Hughbanks, J. Alloys Comp. **358**, 98 (2003).

[15] A. V. Morozkin, J. Alloys. Comp. **353**, L16 (2003).
[16] A. V. Morozkin, J. Alloys Comp. **358**, L9 (2003).
[17] A. V. Morozkin, J. Alloys Comp. **360**, L1 (2003).
[18] A. V. Morozkin, J. Alloys. Comp. **360**, L7 (2003).
[19] L. Chen and J. D. Corbett, Inorg. Chem. **43**, 436 (2004).
[20] A. G. Bolotaev, A. L. Koroliuk, A. V. Morozkin, and V. N. Nikiforov, J. Alloys Comp. **373**, L1 (2004).
[21] A. V. Morozkin, V. N. Nikiforov, and B. Malaman, J. Alloys Comp. **393**, L9 (2005).
[22] A. V. Morozkin, R. Nirmala, and S. K. Malik, J. Alloys Comp. **394**, 75 (2005).
[23] G. Cai, J. Zhang, W. He, P. Qin, and L. Zeng, J. Alloys Comp. **421**, 42 (2006).
[24] A. V. Morozkin, Y. Mozharivskyj, V. Svitlyk, R. Nirmala, O. Isnard, P. Manfrinetti, A. Provino, C. Ritter, J. Solid State Chem. **183**, 1314 (2010).
[25] A. V. Morozkin, R. Nirmala, and S. K. Malik, Internetallics **19**, 1250 (2011).
[26] A. V. Morozkin, V. N. Nikiforov, and B. Malaman, J. Alloys Comp. **393**, L6 (2005).
[27] A. V. Morozkin, O. Isnard, P. Henry, P. Manfrinetti, R. Nirmala, and S. K. Malik, J. Alloys Comp. **450**, 62 (2008).
[28] A. V. Morozkin, O. Isnard, P. Manfrinetti, A. Provino, C. Ritter, R. Nirmala, and S. K. Malik, J. Alloys Comp. **498**, 13 (2010).
[29] A. V. Morozkin, Y. Mozharivskyj, V. Svitlyk, R. Nirmala, and A. K. Nigam, J. Solid State Chem. **183**, 3039 (2010).
[30] A. Herrero, A. Oleaga, A. Salazar, A. V. Garshev, V. O. Yapaskurt, A. V. Morozkin, J. Alloys Comp. **821**, 153198 (2020).
[31] A. Oleaga, A. Herrero, A. Salazar, A. V. Garshev, V. O. Yapaskurt, and A. V. Morozkin, J. Alloys Comp. **843**, 155937 (2020).
[32] A. Herrero, A. Oleaga, I. R. Aseguinolaza, A. J. Garcia-Adeva, E. Apiñaniz, A. V. Garshev, V. O. Yapaskurt, A.V. Morozkin, J. Alloys Comp. **890**, 161849 (2021).
[33] V. Petříček, M. Dušek, and L. Palatinus, Z. Kristallogr. Cryst. Mater. **229**, 345 (2014).
[34] A. M. Clogston, Phys. Rev. Lett. **9**, 266 (1962).
[35] A. Inoue, K. Matsuzaki, T. Masumoto, and H. S. Chen, J. Mater. Sci, **21**, 1258 (1986).

*email: yuchi@issp.u-tokyo.ac.jp
†email: yokamoto@issp.u-tokyo.ac.jp

**Supplementary Note 1. SEM images of a polycrystalline sample of $Zr_6RuBi_2$ (sample #1).**

Enlarged view of Fig. 1(b), which shows SEM image of a polycrystalline sample of $Zr_6RuBi_2$ (sample #1), and the same view with increased contrast are shown in Supplementary Figs. 1(a) and 1(b), respectively.

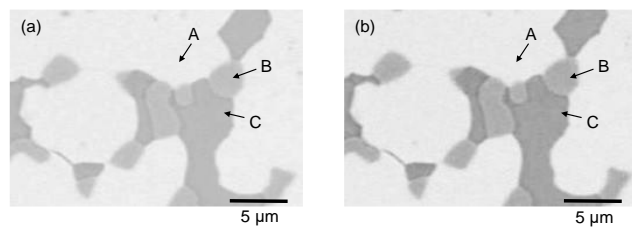

Supplementary Figure 1. Enlarged view of Fig. 1(b) in the main text (a) and that with increased contrast (b).

**Supplementary Note 2. Details on the powder XRD measurements and crystallographic parameters of $Zr_6RuBi_2$ (sample #1) and $Zr_6FeBi_2$.**

$Zr_6MTe_2$ samples have strong preferred orientation, making them difficult to accurately acquire the intensity of each diffraction peak in the XRD data using the Bragg-Brentano method. Therefore, Debye-Scherrer method was used for $Zr_6RuBi_2$ (sample #1) and $Zr_6FeBi_2$, and then one-dimensional data from all directions were converted to reduce the effect of preferred orientation. In a Debye-Scherrer optical system, Mo Kα radiation is generally used, where X-rays can transmit through the sample. On the other hand, since the energy of Mo Kα radiation is higher than that of the L-edges of bismuth, the fluorescence X-rays appear as a strong background in the diffraction data. This increases the overall intensity of the diffraction data, resulting in the larger $R_{exp}$, which is used to evaluate the statistical accuracy of the data. The $S$ values shown in Figs. 1(d) and 1(e) are evaluated using the formula $S = R_{wp}/R_{exp}$, yielding the small $S$ values. In addition, the degree of X-ray absorption strongly correlates with the temperature factor parameters ($U_{eq}$). In the present cases, strong X-ray absorption causes the loss of quantitativity of the values of $U_{eq}$, resulting in a small $U_{eq}$ for Fe in $Zr_6FeBi_2$.

Crystallographic parameters of $Zr_6RuBi_2$ and $Zr_6FeBi_2$ determined by the Rietveld analysis on the powder XRD data are listed in Supplementary Table I. The volume ratio of each phase, i.e., that of $Zr_6RuBi_2$, ZrRu, and Zr for M = Ru and $Zr_6FeBi_2$ and Zr for M = Fe in the Rietveld analysis was 80:12:8 and 90:10, respectively.

Supplementary Table I. Crystallographic parameters for $Zr_6RuBi_2$ and $Zr_6FeBi_2$ determined by the Rietveld analysis of powder XRD data. The space group is $P\text{-}62m$. The lattice parameters are $a = 7.9181(3)$ Å, $c = 3.6847(2)$ Å, and $V = 200.064(17)$ Å$^3$ for $Zr_6RuBi_2$ and $a = 7.8629(4)$ Å, $c = 3.7174(3)$ Å, and $V = 199.04(2)$ Å$^3$ for $Zr_6FeBi_2$.

| Phase | Atom | | $x$ | $y$ | $z$ | $g$ | $U_{eq}$ (Å$^2$) |
|---|---|---|---|---|---|---|---|
| $Zr_6RuBi_2$ | Zr1 | $3g$ | 0.2455(4) | 0 | 1/2 | 1 | 0.0178(13) |
| | Zr2 | $3f$ | 0.6053(4) | 0 | 0 | 1 | 0.0149(14) |
| | Ru1 | $1a$ | 0 | 0 | 0 | 1 | 0.0152(13) |
| | Bi1 | $2d$ | 1/3 | 2/3 | 1/2 | 1 | 0.0152(6) |
| $Zr_6FeBi_2$ | Zr1 | $3g$ | 0.2388(4) | 0 | 1/2 | 1 | 0.0293(15) |
| | Zr2 | $3f$ | 0.6060(4) | 0 | 0 | 1 | 0.0256(16) |
| | Fe1 | $1a$ | 0 | 0 | 0 | 1 | 0.0014(18) |
| | Bi1 | $2d$ | 1/3 | 2/3 | 1/2 | 1 | 0.0219(7) |

**Supplementary Note 3. Characterization and Physical Properties of $Zr_6RuBi_2$ (sample #2).**

An SEM image and powder XRD pattern of a polycrystalline sample of $Zr_6RuBi_2$ (sample #2) prepared under different conditions from that for sample #1 are shown in Supplementary Fig. 2. The synthesis procedure is described in the main text. The heat capacity data for $Zr_6RuBi_2$ (sample #2) are shown in Supplementary Fig. 3.

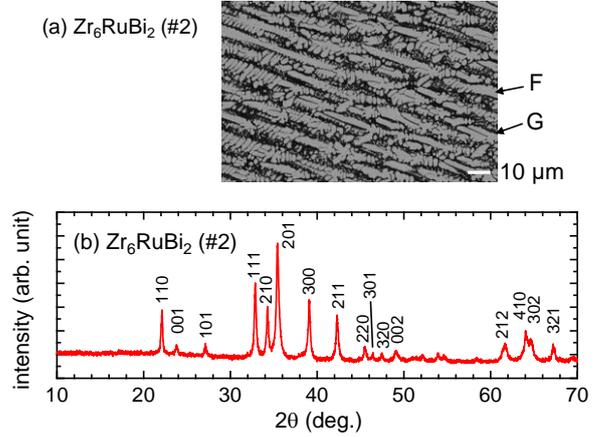

Supplementary Figure 2. (a) SEM image of $Zr_6RuBi_2$ (sample #2). (b) Powder diffraction pattern of $Zr_6RuBi_2$ (sample #2) taken at room temperature using MiniFlex (Cu Kα).

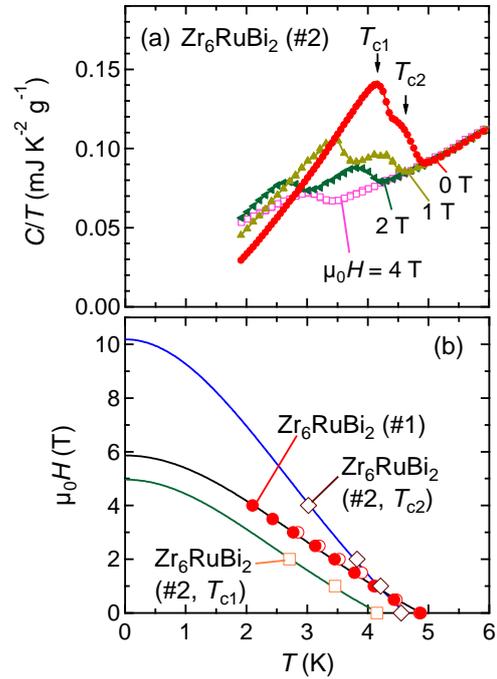

Supplementary Figure 3. (a) Temperature dependence of heat capacity divided by temperature of a polycrystalline sample of $Zr_6RuBi_2$ (sample #2) measured at various magnetic fields. (b) Magnetic field versus $T_{c1}$ and $T_{c2}$ for $Zr_6RuBi_2$ (sample #2). The solid curves show fitting results of $T_{c1}$ and $T_{c2}$ of $Zr_6RuBi_2$ (sample #2) to the Ginzburg-Landau formula. Data for $Zr_6RuBi_2$ (sample #1) shown in Fig. 5 are also included here.